# A Wavelet-Based Digital Watermarking for Video

A.Essaouabi and F.regragui
Department of physics, LIMIARF
Laboratory,
Faculty of Sciences Mohammed V University
Rabat, Morocco

E.Ibnelhaj
Image laboratory
National Institute of Posts and Telecommunications
Rabat, Morocco

*Abstract—* A novel video watermarking system operating in the three-dimensional wavelet transform is here presented. Specifically the video sequence is partitioned into spatio-temporal units and the single shots are projected onto the 3D wavelet domain. First a gray- scale watermark image is decomposed into a series of bitplanes that are preprocessed with a random location matrix. After that the preprocessed bitplanes are adaptively spread spectrum and added in 3D wavelet coefficients of the video shot. Our video watermarking algorithm is robust against the attacks of frame dropping, averaging and swapping. Furthermore, it allows blind retrieval of embedded watermark which does not need the original video and the watermark is perceptually invisible. The algorithm design, evaluation, and experimentation of the proposed scheme are described in this paper.

***Keywords-component;*** *video watermarking; security; copyright protection; wavelet transform*

## I. INTRODUCTION

We have seen an explosion of data change in the Internet and the extensive use of digital media. Consequently, digital data owners can transfer multimedia documents across the Internet easily. Therefore, there is an increase in the concern over copyright protection of digital content [1, 2, 3]. In the early days, encryption and control access techniques were employed to protect the ownership of media. They do not, however protect against unauthorized copying after the media have been successfully transmitted and decrypted. Recently, the watermark techniques are utilized to maintain the copyright [4, 5, 6].

Digital watermarking, one of the popular approaches considered as a tool for providing the copyright protection, is a technique based on embedding a specific mark or signature into the digital products, it has focused on still images for a long time but nowadays this trend seems to vanish. More and more watermarking algorithms are proposed for other multimedia data and in particular for video content. However, even if watermarking still images and video is a similar problem, it is not identical. New problems, new challenges show up and have to be addressed. Watermarking digital video introduces some issues that generally do not have a counterpart in images and audio. Due to large amounts of data and inherent redundancy between frames, video signals are highly susceptible to pirate attacks, including frame averaging, frame dropping, frame swapping, collusion, statistical analysis, etc. Many of these attacks may be accomplished with little or no damage to the video signal. However, the watermark may be adversely affected. Scenes must be embedded with a consistent and reliable watermark that survives such pirate attacks. Applying an identical watermark to each frame in the video leads to problems of maintaining statistical invisibility. Applying independent watermarks to each frame also is a problem. Regions in each video frame with little or no motion remain the same frame after frame. Motionless regions in successive video frames may be statistically compared or averaged to remove independent watermarks [7][8]. In order to solve such problems, many algorithms based on 3D wavelet have been adopted but most of them use the binary image as watermark.

In this paper we propose a new blind watermarking scheme based on 3D wavelet transform and video scene segmentation [8][9]. First By still image decomposition technique a gray- scale watermark image is decomposed into a series of bitplanes which are correlative with each other and preprocessed with a random location matrix. After that the preprocessed bitplanes are adaptively spread spectrum and added in 3D wavelet coefficients of the video shot. As the 1-D multiresolution temporal representation of the video is only for the temporal axis of the video, each frame along spatial axis is decomposed into 2D discrete wavelet multiresolution representations for watermarking the spatial detail of the frame as well as the motion and motionless regions of the video.

Experimental results show that the proposed techniques are robust enough against frame dropping, averaging and MPEG lossy compression.

The rest of this paper is organized as follows: in section II we will explain the decomposition procedure of watermark image and video. Section III will describe the basic functionalities of watermarking embedding and extraction procedure. Finally, section IV will give the simulations results and section V will give the conclusion.

## II. DECOMPOSITION OF THE WATERMARK IMAGE AND VIDEO

### A. Watermark process

The watermark gray scale image $W(i,j)$ is decomposed into 8 bitplanes for watermarking [10]. For robustness to the common picture-cropping processing, a fast two dimensional pseudo-random number traversing method is used to permute each bitplane of the watermark image to disperse its spatial location for the sake of spreading the watermarking information (first key). Finally each bitplane is changed into a pseudo random matrix $W^d_k$ by the disorder processing (second key). $W^d_k$ is a serie of binary image with value 1 and -1.

### B. Decomposition of video.

For watermarking the motion content of video, we decompose the video sequence into multiresolution temporal representation with a 2-band or 3-band perfect reconstruction filter bank by 1-D Discrete Wavelet Transform (DWT) along the temporal axis of the video. To enhance the robustness against the attack on the identical watermark for each frame, the video sequence is broken into scenes and the length of the 1-D DWT depends on the length of each scene. Let N be the length of a video scene, $F^k$ be the k-th frame in a video scene and $WF^k$ be the k-th wavelet coefficient frame. The Wavelet frames are ordered from lowest frequency to highest frequency i.e, $WF^0$ is a DC frame. The procedure of multiresolution temporal representation is shown in Fig.1.

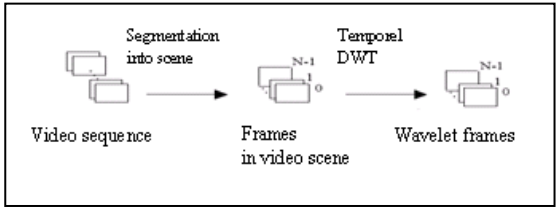

Figure 1.  Procedure of multiresolution temporal representation

The multiresolution temporal representation mentioned above is only along the temporal axis of the video. The robustness of spatial watermarking for each frame (especially for I-frame) should be considered in addition for the sake of surviving MPEG video lossy compression. Hence, the wavelet coefficient frame $WF^k$ is decomposed into multiresolution representation by the 2D discrete Wavelet transform 2DDWT. Fig.2 shows the three-scale discrete wavelet transform with 3 levels using Haar filter. $R_k$ denote the 3D wavelet coefficient frames.

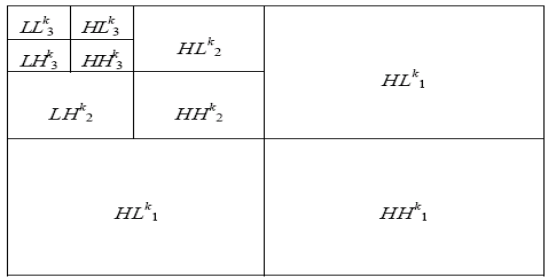

Figure 2.  Three-scale wavelet decomposition with three levels of the k-th wavelet coefficient frame in a video

### III. VIDEO EMBEDDING AND EXTRACTING PROCEDURE

Fig.3 shows the watermarking embedding procedure. Assume that original video is a series of gray-level images of size (352x288) and the watermark image is a 8-bit-grayscale image of size 42x42.

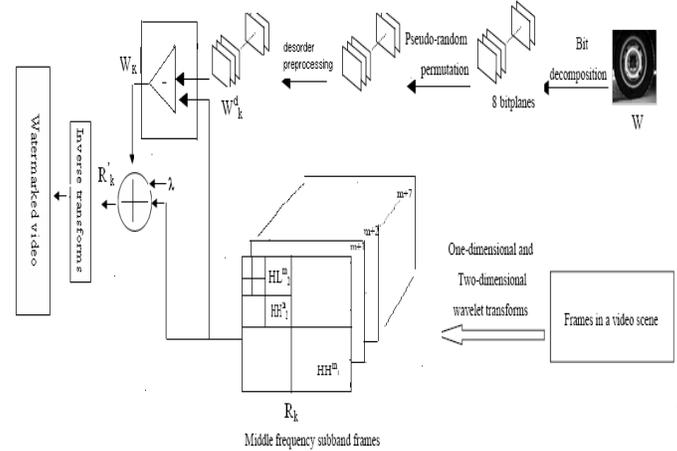

Figure 3.  Digital watermarking embedding scheme diagram

### A. Watermark embedding procedure

The main steps of the digital watermark embedding process are given as follows:

1) **Video segmentation**: the host video is segmented into scene shots, and then some shots are selected by randomly for embedding watermark, then for each scene selected scene shots, the following steps are repeated.

2) The shot is projected by 1-D DWT and 2-D DWT into multiresolution representation of three levels. Denote $R_k$ the 3D wavelet coefficient frames.

3) Each bitplane is adaptively spread spectrum and embedded in each original wavelet coefficient frame (subband LH3). Hence there are 8 original wavelet coefficient frames are watermarked. For each pixel (i,j) of the selected area in $R_K$ (k=1,2,..8), the value is compared with the max of its eight neighbors, t denote the max of its neighbours. Watermark is embedded by changing the corresponding coefficient value as shown Eq 1.

$$R'_K(i,j)= R_K(i,j)+\alpha\, W_K(i,j)\, R_K(i,j) \qquad (1)$$

Where α is an intensity factor, $R'_K$ is the watermarked 3D-DWT coefficient frames, $W_K$ (k=1,2…8) is the spread spectrum watermark image sequence which is the third key of our video watermarking scheme as shown in Eq 2 and Fig.4.

$$W_k(i,j) = \begin{cases} 1 & \text{if } t > R_k(i,j) \text{ and } W^d_k(i,j)=1 \\ & \text{Or } t < R_k(i,j) \text{ and } W^d_k(i,j)=-1 \\ -1 & \text{else} \end{cases} \quad (2)$$

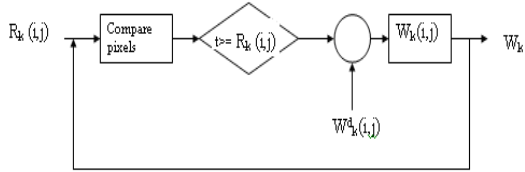

Figure 4. The detail of the watermark embedding

4) By inversing the watermarked 2D-DWT and 1-D DWT wavelet coefficient frames, we obtain the watermarked video.

*B. Watermark extracting procedure*

1) We first parse the watermarked video into shots with the same algorithm as watermark embedding, and then the 3D wavelet transform is performed on each selected test video shot, for each wavelet coefficient frame $R'_K$ (k=1,2,…n).
2) For each pixel in $R'_K(i,j)$, its value is compared with max of its eight neighbors. t' denotes the max value of its eight neighbors to extract the corresponding bitplane. As shown in Fig.5 and Eq.3

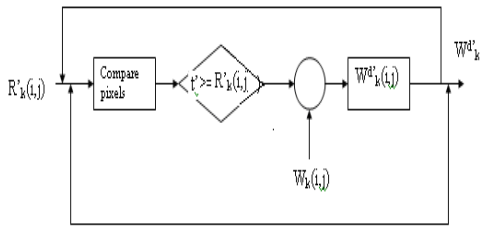

Figure 5. The detail of the watermark detecting

$$W^d_k(i,j) = \begin{cases} 1 & \text{if } t > R'_k(i,j) \text{ and } W_k(i,j) = 1 \\ & \text{Or } t < R'_k(i,j) \text{ and } W_k(i,j) = -1 \\ -1 & \text{else} \end{cases} \quad (3)$$

3) The preprocessing and the pseudo-random permutation is reversed according to the predefined pseudo-random order for these bitplanes
4) By composing these bitplanes into the gray-level image G0 the extracted watermark is reconstructed

IV. EXPERIMENTAL RESULTS

The "foreman" and "Stefan" sequences with 100 frame long (about 4 seconds) and 352x288 pixels per frame as shown in (Fig 6-a et b) were used in our experiments. The image tire (watermark) 42x42 that we used in our experiments is shown in (Fig 6-c).

The corresponding experiment results for various possible attacks such as frame dropping, frame averaging, frame swapping, and MPEG compression are shown as follow section, in the other hand a similarity measurement of the extracted and the referenced watermarks is used for objective judgment of the extraction fidelity and it is defined as:

$$NC = \frac{\sum_i \sum_j W(i,j)W'(i,j)}{\sum_i \sum_j [W(i,j)]^2} \quad (4)$$

which is the cross-correlation normalized by the reference watermark energy to give unity as the peak correlation. We will use this measurement to evaluate our scheme in our experiment.

Peak signal-to-noise ratio (PSNR), a common image quality metric, is defined as:

$$PSNR = 20\log\left(255 / \sqrt{SNR}\right) \quad (5)$$

The signal-to-noise ratio (SNR) is computed between the original and watermarked frame.

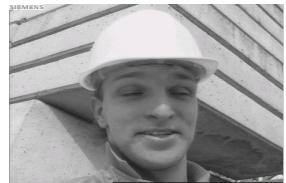 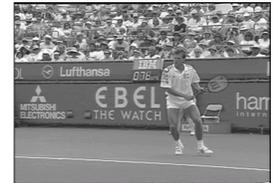

(a) Foreman scene (b) Stefan scene

(c) Original watermark

Figure 6. Two scenes original and the watermark image in the experiment

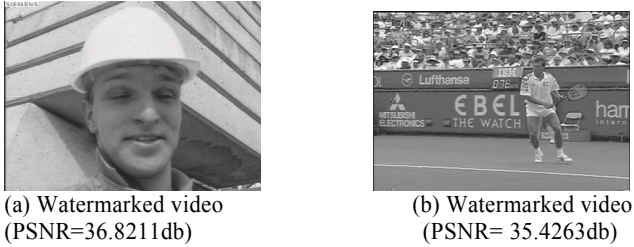

(a) Watermarked video  (b) Watermarked video
(PSNR=36.8211db)   (PSNR= 35.4263db)

Figure 7. The watermarked scenes

(a) Extracted watermark from foreman scene NC(0.9736)

(b) Extracted watermark from stefan scene NC(0.9587)

Figure 8. The extracted watermark from each scene

*A. Frame dropping attack*

There is a little change between frames in shot .so the frame dropping which are some frames (even index frame)are removed from the video shot and replaced by corresponding original frames is used as an effective video watermark attack.
The experimental result is plotted in Fig.9

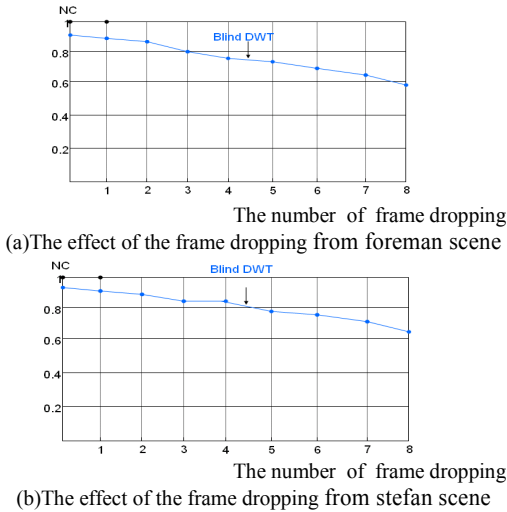

(a)The effect of the frame dropping from foreman scene

(b)The effect of the frame dropping from stefan scene

Figure 9. NC values under frame dropping. From the experiment, we found that our scheme achieves better performance

Frame averaging is also a significant video watermarking attack that will remove dynamic composition of the video watermarked so in our experiment we use the average of current frame and its two nearest neighbors to replace the curren $k=2,3,4,\ldots\ldots n-1$,
the corresponding results are presented in Fig.10

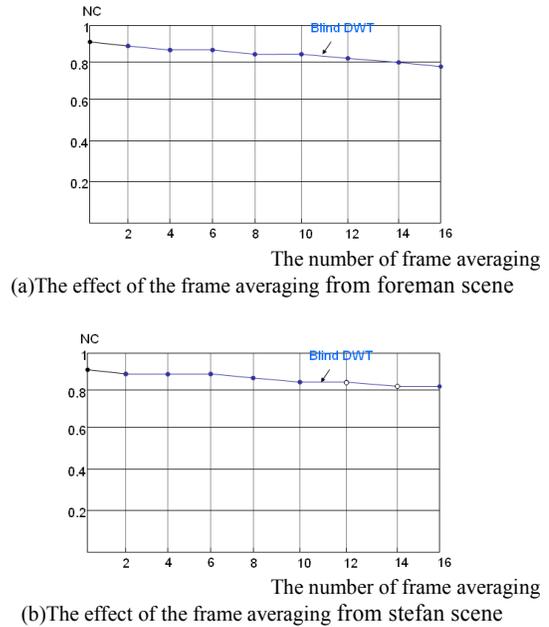

(a)The effect of the frame averaging from foreman scene

(b)The effect of the frame averaging from stefan scene

Figure 10. NC values under statistical averaging. It is found that the proposed scheme can resist to statistical averaging quite well.

*B. Frame swapping attack*

Frame swapping can also destroy some dynamic composition of the video watermark.
We define the following swapping mode by $F_K(i,j)= F_{k-1}(i,j)$ $k=1,3,5\ldots\ldots.n-1$
the corresponding results are presented in Fig.11

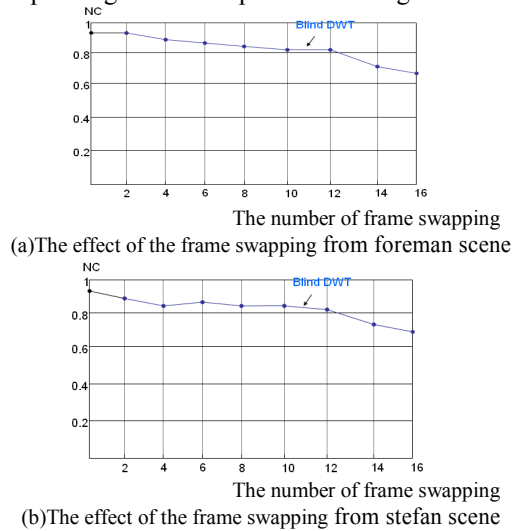

(a)The effect of the frame swapping from foreman scene

(b)The effect of the frame swapping from stefan scene

Figure 11. NC values under frame swapping. From the experiment, we found that our scheme achieves better performance.

*C. MPEG compression*

MPEG compression is one of the basic attacks to video watermark. The video watermarking scheme should robust against it.fig.12 shows the extracted watermark from foreman scene after MPEG2 compression.

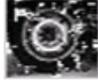

Figure 12. Extracted watermark after MPEG2 compression

## V. CONCLUSION

This paper proposes an innovative blind video watermarking scheme in the 3D wavelet transform using a gray scale image as a watermark.

The process of this video watermarking scheme, including watermark preprocessing, video preprocessing, watermark embedding, and watermark detection, is described in detail. Experiments are performed to demonstrate that our scheme is robust against attacks by frame dropping, frame averaging, and lossy compression.